\documentclass[fleqn,usenatbib]{mnras}

\usepackage{newtxtext,newtxmath}
\usepackage{comment}

\usepackage[T1]{fontenc}

\DeclareRobustCommand{\VAN}[3]{#2}
\let\VANthebibliography\thebibliography
\def\thebibliography{\DeclareRobustCommand{\VAN}[3]{##3}\VANthebibliography}

\usepackage{graphicx}	
\usepackage{amssymb}	
\usepackage{amsmath}	

\def\aj{AJ}             	
\def\araa{ARA\&A}       	
\def\apj{ApJ}           	
\def\apjl{ApJ}          	
\def\apjs{ApJS}         	
\def\aap{A\&A}          	
\def\mnras{MNRAS}       	

\def\H2{H$_{\rm 2}$}

\title[The CO Universe]{The CO universe: Modelling CO emission and H$_{\rm 2}$ abundance in cosmological galaxy formation simulations}

\author[S. Inoue, N. Yoshida \& H. Yajima]{
{Shigeki Inoue$^{1,2,3,4}$\thanks{E-mail: shigeki.inoue@nao.ac.jp}, Naoki Yoshida$^{3,4,5}$ \& Hidenobu Yajima$^{1}$}
\\
$^{1}$Center for Computational Sciences, University of Tsukuba, Ten-nodai, 1-1-1 Tsukuba, Ibaraki 305-8577, Japan\\
$^{2}$Chile Observatory, National Astronomical Observatory of Japan, Mitaka, Tokyo 181-8588, Japan\\
$^{3}$Kavli Institute for the Physics and Mathematics of the Universe (WPI), UTIAS, The University of Tokyo, Chiba 277-8583, Japan\\
$^{4}$Department of Physics, School of Science, The University of Tokyo, Bunkyo, Tokyo 113-0033, Japan\\
$^{5}$Research Center for the Early Universe, School of Science, The University of Tokyo, Bunkyo, Tokyo 113-0033, Japan
}

\date{Accepted XXX. Received YYY; in original form ZZZ}

\pubyear{2020}

\begin{document}
\label{firstpage}
\pagerange{\pageref{firstpage}--\pageref{lastpage}}
\maketitle

\begin{abstract}
We devise a physical model of formation and distribution of molecular gas clouds in galaxies. We use the model to predict the intensities of rotational transition lines of carbon monoxide (CO) and the molecular hydrogen (H$_{\rm 2}$) abundance. Using the outputs of Illustris-TNG cosmological simulations, we populate molecular gas clouds of unresolved sizes in individual simulated galaxies, where the effect of the interstellar radiation field with dust attenuation is also taken into account. We then use the publicly available code DESPOTIC to compute the CO line luminosities and H$_{\rm 2}$ densities without assuming the CO-to-H$_{\rm 2}$ conversion factor ($\alpha_{\rm CO}$). Our method allows us to study the spatial and kinematic structures traced by CO(1-0) and higher transition lines. We compare the CO luminosities and H$_{\rm 2}$ masses with recent observations of galaxies at low and high redshifts. Our model reproduces well the observed CO-luminosity function and the estimated H$_{\rm 2}$ mass in the local Universe. About ten percent of molecules in the Universe reside in dwarf galaxies with stellar masses lower than $10^9~{\rm M_\odot}$, but the galaxies are generally `CO-dark' and have typically high $\alpha_{\rm CO}$. Our model predicts generally lower CO line luminosities than observations at redshifts $z\gtrsim 1$--$2$. We argue that the difference can be explained by the highly turbulent structure suggested for the high-redshift star-forming galaxies.
\end{abstract}




\section{Introduction}
\label{Intro}
Molecular gas clouds (MGCs) are the birthplaces of stars, and understanding the formation and the abundance of MGCs is important in the study of galaxy formation and evolution. Despite the ubiquitous existence, direct observation of \H2 molecules in the Universe is severely limited. Radiation from \H2 molecules in the cold inter-stellar medium (ISM) is not observable because of the high excitation energies. Rotational transition lines of CO molecules are often used as a proxy for \H2 in the ISM. CO is the second most abundant molecules in the Universe, and a variety of emission lines can be observed in radio to submillimetre bands. 

 The rotational transition line from the state of $J=1$ to $0$, CO(1-0), is thought to trace the surface density of \H2, and the conversion factor $\alpha_{\rm CO}$ between them is often assumed to be constant (see Section \ref{lowz}). However, it is also known that the value of $\alpha_{\rm CO}$ varies depending on locations in a galaxy, galactic types and redshifts \citep[e.g.][]{bwl:13}. Also there are many `CO-dark' MGCs, where CO(1-0) emission is significantly weaker than expected for the estimated \H2 density. The variation of $\alpha_{\rm CO}$ likely reflects differences in the local environments and physical properties among different MGCs, such as their column densities, intensity of far-ultraviolet (FUV) radiation fields, dust and metal abundances \citep[e.g.][see also Section \ref{simandmethod}]{nko:12,lnd:18}.
 
Theoretical studies have been hampered by difficulties associated with the thermal and chemical structure and evolution of MGCs. Since most \H2 molecules are formed on the surface of dust grains, one needs to model the formation, growth and destruction of dust grains that involve various physics and chemistry even at microscopic levels. In addition, as shown in radiative transfer simulations of \citet{gfm:10}, the \H2 abundance can be large in central regions of MGCs that are self-shielded against external radiation fields, which means that calculating molecular abundances needs detailed radiative transfer computations with resolving the small-scale internal structure. The conditions are simpler for CO molecules. Formation of CO does not rely on dust grains, although it needs \H2 \citep[e.g.][]{k:17,gok:20}. The self-shielding effect of CO is less important than \H2. In addition, metallicity gradients in galaxies can enhance the inhomogeneity of the abundance ratios. These facts actually suggest that the abundance ratio between \H2 and CO may not be uniform. Numerical simulations aimed at predicting accurately the molecular abundances and line emissivities are generally required to have extremely high resolutions and to implement molecular chemistry, radiation transfer and dust models; \citet{gfm:10} and \citet{gbs:14} study the convergence with respect to resolutions in their detailed radiative transfer simulations. 

Because formation of molecules does not alter the gravitational assembly nor overall dynamical evolution of galaxies, often post-processing methods have been applied to outputs of simulations in order to compute molecular abundances. A number of studies focus on a single galaxy with sufficiently high resolution to resolve giant molecular clouds \cite[e.g.][]{nko:12,lnd:18,vpf:18,akd:20,krm:20,lgy:20}. These studies treat a small number of galaxies in isolated or in  cosmological simulations and thus statistical quantities such as line luminosity functions cannot be reliably determined. Also, possible redshift evolution has not been fully addressed. 

Large-volume cosmological simulations are indispensable for statistical studies of galaxy {\it populations}, but the typically poor mass resolution achieved to date still hampers us from directly representing individual MGCs. Also it is computationally expensive to perform multi-dimensional radiation transfer for a large number of galaxies even if the mass and spatial resolutions were appropriate. Therefore, previous studies on galaxy populations utilise large-box simulations by employing simple, empirical models or semi-analytic approaches \citep[e.g.][]{ocd:09,lbl:11,pbp:15,srp:20,dcs:20}. \citet{pps:19} apply a post-processing method to the results of a cosmological simulation and of a semi-analytic model based on a fit to \H2 fraction obtained by \citet[][also see \citealt{dsl:19}]{gk:11}. They study in detail the evolution of \H2 mass functions of galaxies from redshift $z=0$ to $5$. The derived \H2 masses are in agreement with observations of local CO-luminosity functions assuming a constant $\alpha_{\rm CO}$. However, their results predict \H2 masses that are significantly smaller than high-redshift CO observations if they adopt the same $\alpha_{\rm CO}$. As a possible reason, they argue that $\alpha_{\rm CO}$ may decrease with redshift on average in the observed galaxies.

It seems that theoretical studies tend to focus on \H2 as it is more directly linked to star formation (SF). Unfortunately, \H2 is not readily observable, and thus modelling CO formation is necessary to make direct comparison with observations. It is also important to study higher rotational transition. Although high-redshift observations of CO lines are currently performed by Atacama Large Millimeter/submillimeter Array (ALMA), redshifted CO(1-0) lines from distant galaxies shift to outside the observable wavelength range of ALMA. The current high-redshift CO observations, therefore, rely on high-$J$ transition lines, which are often converted to derive expected CO(1-0) emission with {\it assuming} some certain spectral line energy distribution (see Section \ref{aspecs}). 

In this paper, we propose a physical model of MGC formation and distribution, and compute \H2 density and CO line luminosity in a consistent manner. We apply the post-processing method to the outputs of a large-box cosmological simulation. Our model allows us to calculate intensities of not only CO(1-0) but also high-$J$ lines. We make direct comparison between the simulation and observations without converting CO luminosity to \H2 density, and (dis)agreement between them could help us to understand differences of ISM and cloud properties among galaxies and their evolution with redshift. 

In the rest of the present paper, Section \ref{simandmethod} describes the simulation we utilise, our modellings for gas clouds and related parameters. Section \ref{chemrt} gives our computations to obtain molecular abundances and line intensities. Section \ref{results} presents our results and comparison with recent survey observations for nearby and distant galaxies. Section \ref{discuss} discusses agreement and disagreement between our results and observations. There, we address possible redshift-evolution of molecular clouds in the Universe. Section \ref{conclusions} summarises our findings.

\section{Modelling Methods}
\label{simandmethod}

Our method is based on post-processing a large-volume cosmological simulation. We aim at obtaining the large-scale distribution of galaxy populations as well as resolving spatial and kinematic structures of gas, stars and dark matter in individual galaxies. Unfortunately, even state-of-the-art cosmological simulations do not fully resolve the distribution of stars and gas clouds at length scales of $\lesssim100~{\rm pc}$. We thus resort to applying a physically motivated model to populate the galaxies with MGCs.  In Section \ref{chemrt}, we compute the molecular abundance and atomic/molecular line emissivities using approximate radiation transfer with chemistry for individual MGCs. To this end, we also need to estimate strength of radiation affecting molecular abundances of the gas clouds, and we employ simple models for radiation fields and dust attenuation.

\subsection{Cosmological simulation}
\label{sims}
We utilise the data set of IllustrisTNG simulations \citep{TNG}. The details of the simulations are presented on the IllustrisTNG project web site\footnote{https://www.tng-project.org/} and in related papers such as \citet{TNG}, \citet{wsh:17} and \citet{psn:18}. We specifically use the outputs of TNG100-1 run. The simulation box has a comoving side length of $75~{\rm Mpc}$, and the mass-resolutions for dark matter and gas are $7.5$ and $1.4\times10^6~{\rm M_\odot}$. In the simulation, dense gas cells with $\rho_{\rm cell}>n_{\rm H,SF}=0.1~{\rm cm^{-3}}$ are converted to stellar particles according to a stochastic SF model. Therefore the stellar mass resolution is roughly the same as that of the parent gas cells. 

The star formation rate (SFR) is calculated as
\begin{equation}
\dot{m}_{\rm star} = f_{\rm M} \frac{m_{\rm cell}}{t_{\rm SF}}
\end{equation}
where $m_{\rm cell}$ is a mass of the parent cell, the factor $f_{\rm M}$ is the mass fraction of cold gas  (see Section \ref{cloudmodel}), and $t_{\rm SF}=(G\rho_{\rm cell})^{-1/2}$. The ISM model of  \citet[][see also \citealt{sh:03}]{ykk:97} is adopted to the star-forming gas with $\rho_{\rm cell}>n_{\rm H,SF}$. Type-II supernovae (SNe) are triggered immediately following the SF, and a mechanical feedback model of \citet{sh:03} is adopted to represent stellar feedback effects. Type-Ia SNe and asymptotic giant branch stars eject mass and metals into nearby gas cells. The simulation also implements creation and feedback of black holes \citep{sdh:05} and magnetic fields \citep{pms:14}.

Gravitationally bound structures are identified with the friend-of-friend and {\sc SUBFIND} grouping algorithms \citep[e.g.][]{swt:01}. The total masses and line luminosities are computed for each {\sc SUBFIND} group (galaxy). 

\subsection{Populating molecular gas clouds}
\label{cloudmodel}
\begin{figure}
  \includegraphics[bb=0 0 779 488, width=\hsize]{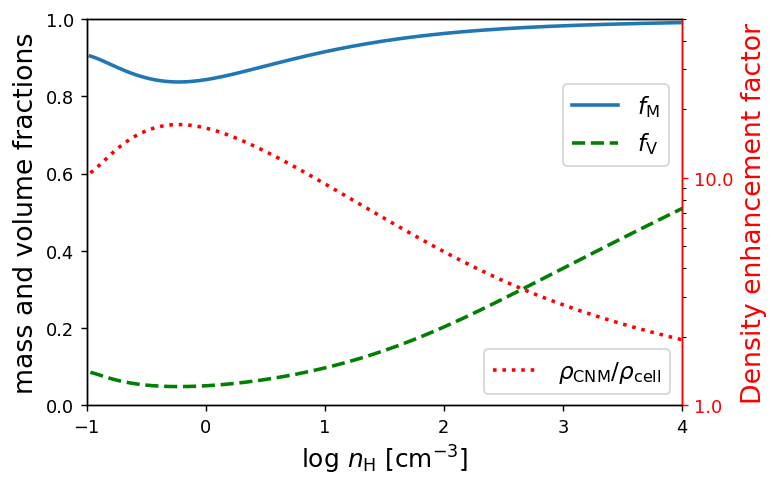}
  \caption{We plot the fractions of CNM as a function of the gas density. \textit{Blue solid line:} the mass fraction of CNM of a gas cell at $z=0$, which is computed by the two-phase ISM model using the same cooling function as that of IllustrisTNG simulation. \textit{Green dashed line:} the volume fraction of CNM, which is computed by equation (\ref{volumefrac}) assuming the density contrast $\phi=100$. \textit{Red dotted line with the right ordinate:} density enhancement factor $\rho_{\rm CNM}/\rho_{\rm cell}=f_{\rm M}/f_{\rm V}$. Note the logarithmic scales.}
  \label{fractions}
\end{figure}
We first need to determine the mass distribution of MGCs and their physical properties. We achieve this by using the quantities of individual gas cells in the simulation as follows. In the IllustrisTNG simulations, the ISM model of \citet{ykk:97} is applied to gas with $\rho_{\rm cell}>n_{\rm H,SF}=0.1~{\rm cm^{-3}}$. The dense gas is assumed to consist of cold and hot phases; the hot gas can cool and be converted to the cold gas by thermal instability, and the cold gas triggers "unresolved" SF and SNe. The SNe evaporate some amount of cold gas back to hot one, and the ejecta is turned into the hot phase. By considering the pressure equilibrium, the two-phase ISM model yields a barotropic equation of state (EOS). Since this effective EOS is significantly harder than the isothermal EOS, the two-phase ISM model in cosmological simulations tend to prevent galactic discs from fragmenting and forming giant clumps \citep{iy:19}. 

The two-phase ISM model computes the mass fraction of the cold-phase gas for each gas cell, $f_{\rm M}$, as a function of density for a given cooling function. Typically $f_{\rm M}\gtrsim0.8$ for all densities above $n_{\rm H,SF}$ and approaches asymptotically to $f_{\rm M}\simeq1$ at high densities (see the blue solid line in Fig. \ref{fractions}). Thus, the cold-phase gas is dominant in mass in star-forming regions.

\citet{wmh:95} show that there exist two equilibrium states for a neutral gas at a given pressure, corresponding to warm and cold neutral media (WNM and CNM). They also find that the density contrast between the two phases is always $\phi\equiv\rho_{\rm CNM}/\rho_{\rm WNM}\sim100$, and hardly depends on the physical properties of the gas nor on external radiation. We thus assume $\phi=100$ in our model. Following \citet{dsl:19} and \citet{pps:19}, we assume that all the gas in cells with $\rho_{\rm cell}>n_{\rm H,SF}$ is neutral (atomic and molecular), whereas the gas in cells with $\rho_{\rm cell}<n_{\rm H,SF}$ is fully or partly ionised and contains no molecules. For the dense gas, the total cell mass is given by $m_{\rm cell}=m_{\rm CNM} + m_{\rm WNM}=\rho_{\rm CNM}V_{\rm CNM} + \rho_{\rm WNM}V_{\rm WNM}$ and volume $V_{\rm cell}=m_{\rm cell}/\rho_{\rm cell}=V_{\rm CNM} + V_{\rm WNM}$, where $V_{\rm WNM}$ and $V_{\rm CNM}$ are the volumes of WNM and CNM within the cell. We assume that the hot and cold phases defined in the two-phase ISM model correspond to the WNM and CNM in \citeauthor{wmh:95}, and compute the density and volume of the CNM. The CNM volume fraction is 
\begin{equation}
f_{\rm V}\equiv\frac{V_{\rm CNM}}{V_{\rm cell}}=\frac{f_{\rm M}}{\left(1-f_{\rm M}\right)\phi+f_{\rm M}},
\label{volumefrac}
\end{equation}
where $f_{\rm M}=m_{\rm CNM}/(m_{\rm CNM} + m_{\rm WNM})$. 
Fig. \ref{fractions} shows $f_{\rm V}$ as a function of density. In spite of the dominance in mass (the blue solid line), CNM occupies a small volume at $\rho_{\rm cell}\lesssim10^2~{\rm cm^{-3}}$ when $\phi=100$. The density enhancement factor is defined as 
\begin{equation}
\frac{\rho_{\rm CNM}}{\rho_{\rm cell}}=\frac{f_{\rm M}}{f_{\rm V}}
\end{equation}
which is shown by the red dotted line in Fig. \ref{fractions}. We assume that molecules are formed only in CNM. 
 
Next, we need to determine the physical size of a MGC. Molecules can be photo-dissociated by the inter-stellar radiation field (ISRF) but the dissociating radiation cannot penetrate deep into the innermost regions if the gas column density is high enough for self- and/or dust-shielding against the ISRF. The degree of the shielding effects is determined by the MGC size and density. Jeans length $\lambda_{\rm J}$ is often used to approximate the characteristic size of a MGC.
We assume that a cloud in the CNM has a radius 
 \begin{equation}
r_{\rm cloud}=\frac{\lambda_{\rm J}}{2}=\frac{1}{2}\sqrt{\frac{\gamma P_{\rm CNM}}{G\rho^2_{\rm CNM}}},
\label{jeans}
\end{equation}
where $G$ is gravitational constant and $\gamma=5/3$
is the adiabatic index. Note that our base model does not consider turbulent nor magnetic pressure (although see discussion Section \ref{discusshighz}). Assuming the pressure equilibrium between CNM and WNM, the pressure of CNM is equal to that of the cell, i.e. $P_{\rm CNM}=P_{\rm cell}$. The mass and the column density of a single cloud are given by
\begin{equation}
m_{\rm cloud}=\rho_{\rm CNM}V_{\rm cloud}=\frac{4\pi}{3}\rho_{\rm CNM}\,r_{\rm cloud}^3
\end{equation}
and
\begin{equation}
\Sigma_{\rm cloud}=\frac{m_{\rm cloud}}{\pi r_{\rm cloud}^2}=\frac{4}{3}\rho_{\rm CNM} \,r_{\rm cloud}.
\label{column}
\end{equation}
The above quantities of $\rho_{\rm CNM}$ and $\Sigma_{\rm cloud}$ are used as parameters when we compute molecular abundances and emissivities in Section \ref{despotic}.

\subsection{Inter-stellar radiation field}
\label{isrf}
Molecular line emission is powered by an internal or external radiation. To estimate the ISRF strength, we use a simple parametric model and calibrate it to match available observations. FUV radiation is most relevant for photo-dissociation of molecules, which is mainly emitted from young massive stars. We assume that the unattenuated ISRF intensity, $\chi_{\rm int}$, scales with the total SFR within a galaxy as 
\begin{equation}
\chi_{\rm int}=\chi_\odot\frac{\dot{M}_{\rm star}}{1{\rm M_{\odot}~{\rm yr^{-1}}}},
\label{intISRF}
\end{equation}
where $\chi_\odot$ is the ISRF intensity in the solar neighbourhood. 

To calculate the dust-attenuated ISRF, we estimate the amount of dust within a radius characterised by the SFR distribution since FUV radiation is primarily contributed by local SF in the galaxy. The dust column density is approximated to be 
\begin{equation}
\Sigma_{\rm dust}=\frac{f_{\rm dust}M_{\rm metal}(<r_{\rm SFR})}{\pi r_{\rm SFR}^2},
\label{dustcolumn}
\end{equation}
where $r_{\rm SFR}$ is the three-dimensional radius within which half the total SFR of the galaxy is enclosed, $M_{\rm metal}(<r_{\rm SFR})$ is the total metal mass within $r_{\rm SFR}$, and we assume a constant dust-to-metal fraction to be $f_{\rm dust}=0.3$ in our fiducial case. Assuming the typical size and solid density of a dust grain to be $a=0.01~\mu{\rm m}$ and  $s=3.0~{\rm g~cm^{-3}}$, the optical depth is approximately given by
\begin{equation}
\tau=\frac{3\Sigma_{\rm dust}}{4as}.
\label{dusttau}
\end{equation}
Note that the dust attenuation we consider here is different from dust-shielding within a MGC. Later in Section \ref{despotic}, we discuss the dust-shielding effect on molecular emission.

In addition to the ISRF, there may exist external radiation such as the cosmic background radiation. Its intensity $\chi_{\rm ext}$ is uniform but may vary with redshift. We follow the analytic model of \citet{phh:19} and set the intensity at wavelength of $1000~{\rm \AA}$ as $\log(\chi_{\rm ext}/\chi_\odot)\simeq-3.0$ at $z=0$. The background intensity monotonically increases to $-1.0$ at $z=6$.\footnote{The values of $\log(\chi_{\rm ext}/\chi_\odot)\simeq-2.0$, $-1.5$, $-1.3$, $-1.1$ and  $-1.0$ at redshifts $z=1$, $2$, $3$, $4$ and $5$, respectively.} 

Finally, the radiation intensity in a galaxy is modelled as 
\begin{equation}
\chi_{\rm cloud} = \chi_{\rm int}\exp\left(-\tau\right) + \chi_{\rm ext},
\label{ISRF}
\end{equation}
and this is another parameter of the molecular computations in Section \ref{despotic}. Our model assumes a constant $\chi_{\rm cloud}$ to all the MGCs in a single galaxy. This approximation may not represent accurately local variations of molecular abundances in, for instance, spiral arms and inter-arm regions in a galaxy. We focus on the statistical quantities such as line luminosity functions for populations of galaxies in the present paper. Detailed radiative transfer within clumpy galaxies will be a subject of our future study.

\section{Computations for molecular abundances and emission}
\label{chemrt}
\subsection{Creating look-up tables}
\label{despotic}

We use the radiation transfer code {\sc DESPOTIC} \citep[Derive the Energetics and SPectra of Optically Thick Interstellar Clouds, see][]{despotic:14} that can compute the abundances of various chemical species and can predict atomic/molecular line emission from a gas cloud. Here we focus on \H2 and CO molecules. {\sc DESPOTIC} employs a spherical one-zone cloud model, and includes carbon chemistry network as well as various physical processes for cooling and heating of cold ISM. For a specified parameter set that describes the physical properties of a gas cloud, the code self-consistently calculates a thermal and chemical equilibrium state. It then returns the gas and dust temperatures, species' abundances and line emissivities.

The basic parameters to be input to {\sc DESPOTIC} are: volume and column densities of a cloud and the ISRF intensity. 
For these, we use the values of $\rho_{\rm CNM}$, $\Sigma_{\rm cloud}$ and $\chi_{\rm cloud}$ derived in the previous sections. Non-thermal velocity dispersion $\sigma$ of the cloud is calculated with the assumption of a marginally bound state by gravity, where the virial parameter $\alpha_{\rm vir}\equiv5\sigma^2r_{\rm cloud}/(Gm_{\rm cloud})=1$ \citep[][and references therein]{hd:15}, 
\begin{equation}
\sigma = \sqrt{\frac{3\pi G\Sigma_{\rm cloud}^2}{20\rho_{\rm CNM}}}.
\label{sigmaNT}
\end{equation}
Although considering non-thermal (turbulent) motions in DESPOTIC might appear inconsistent with the assumption made when calculating $r_{\rm cloud}$ (Section \ref{cloudmodel}), we adopt the above equation by noting that MGCs can be highly turbulent after gravitational contraction. \citet{fk:12} demonstrate that efficient star formation is driven by compressive turbulence \citep[see also][]{f:18}. We also note that the effect of varying $r_{\rm cloud}$ shall be discussed in Section \ref{discusshighz}. The ionization rate due to hard X-ray photons and cosmic rays is set to $\xi=10^{-17}\chi_{\rm cloud}~{\rm s^{-1}}$ per H nucleus. We assume the dust abundance in a cloud to be proportional to its metallicity $Z_{\rm cloud}$. With the efficient mixing approximation within a gas cell, we consider $Z_{\rm cloud}=Z_{\rm cell}$ but impose the minimum metallicity of $10^{-3}Z_\odot$ set by Population III stars \citep{kkm:12}, where $Z_\odot$ is the solar metallicity. The dust abundance of the cloud is given as
\begin{equation}
D_{\rm cloud}=D_{\rm MW}\frac{Z_{\rm cloud}}{Z_\odot},
\label{Dcloud}
\end{equation}
where $D_{\rm MW}$ is the Milky Way dust abundance. We adopt the total abundance of ${\rm [C/H]} = 2\times10^{-4}Z_{\rm cloud}/Z_\odot$, ${\rm [O/H]} = 4\times10^{-4}Z_{\rm cloud}/Z_\odot$ and ${\rm [M/H]} = 2\times10^{-7}Z_{\rm cloud}/Z_\odot$ for C, O and the other refractory metals (M), respectively.\footnote{Although these values can be directly read from the snapshot data of Illustris-TNG, we assume the simple scaling with metallicity to reduce the number of input parameters. The actual abundances in the simulation do not significantly deviate from the scaling relations.} These abundances are consistent with the solar value for $Z_{\rm cloud}=Z_\odot$ \citep{d:11}. For dust grains, we set three cross sections per H nucleus: one for thermal radiation $\sigma_{\rm 10}=2\times10^{-25}~{\rm cm^{-2}}$ at 10 K, one for photoelectric heating $\sigma_{\rm PE}=10^{-21}~{\rm cm^{-2}}$ and yet another one for ISRF $\sigma_{\rm ISRF}=3\times10^{-22}~{\rm cm^{-2}}$. The dust-gas coupling coefficient is set to $\alpha_{\rm gd}=3.2\times10^{-34}$. The spectral index for dust thermal radiation is $\beta=2.0$. Finally, the cosmic microwave background temperature is $T_{\rm CMB}=2.73~{\rm K}$ at $z=0$.

With these settings, {\sc DESPOTIC} needs five parameters: $\rho_{\rm CNM}$, $\Sigma_{\rm cloud}$, $\chi_{\rm cloud}$, $Z_{\rm cloud}$ and $z$. Since $\Sigma_{\rm cloud}$ is a function of $\rho_{\rm CNM}$ in the two-phase ISM model used in Illustris-TNG and the cloud size $r_{\rm cloud}=\lambda_{\rm J}/2$ depends only on $\rho_{\rm CNM}$ at a given $z$,
we generate a look-up table of \H2 fractions $f_{\rm H_2}$ and line emissivities $W_{\rm CO}$ as a function of $\rho_{\rm CNM}$, $\chi_{\rm cloud}$ and $Z_{\rm cloud}$ at each output epoch (redshift). The parameter space covers the values of the three quantities with $30$, $10$ and $10$ grids for $\rho_{\rm CNM}$, $\chi_{\rm cloud}$ and $Z_{\rm cloud}$ in logarithmic spacing.

\subsection{Integrating the clouds}
\label{integclouds}
Using the look-up table, we compute CO line intensities and \H2 abundance for each gas cell according to the procedures described in Section \ref{cloudmodel}. We do not consider inter-galactic components that are not gravitationally bound to any galaxies, because such a warm/hot, diffuse gas is highly ionised and contains little molecules.

\begin{figure}
  \includegraphics[bb=0 0 616 381, width=\hsize]{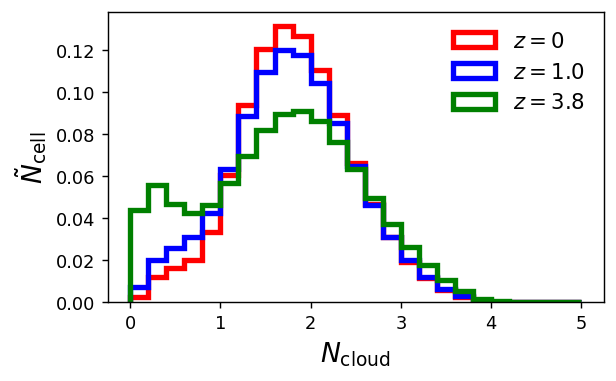}
  \caption{Normalised histograms of gas cells as functions of $N_{\rm cloud}$ computed by equation (\ref{n_clouds}) in the snapshots at redshifts $z=0$, $1.0$ and $3.8$.}
  \label{Ncellhist}
\end{figure}
\begin{figure*}
  \includegraphics[bb=0 0 1534 671, width=\hsize]{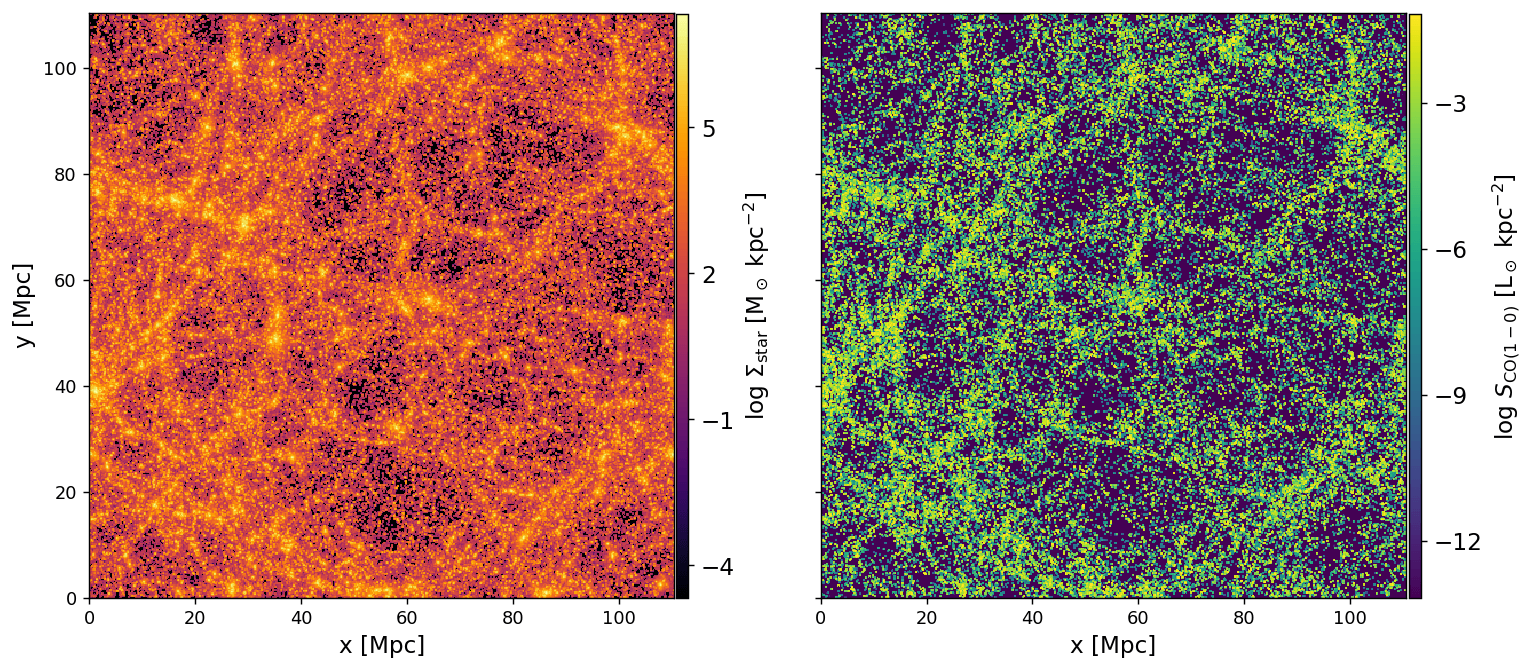}
  \includegraphics[bb=0 0 1756 1317, width=\hsize]{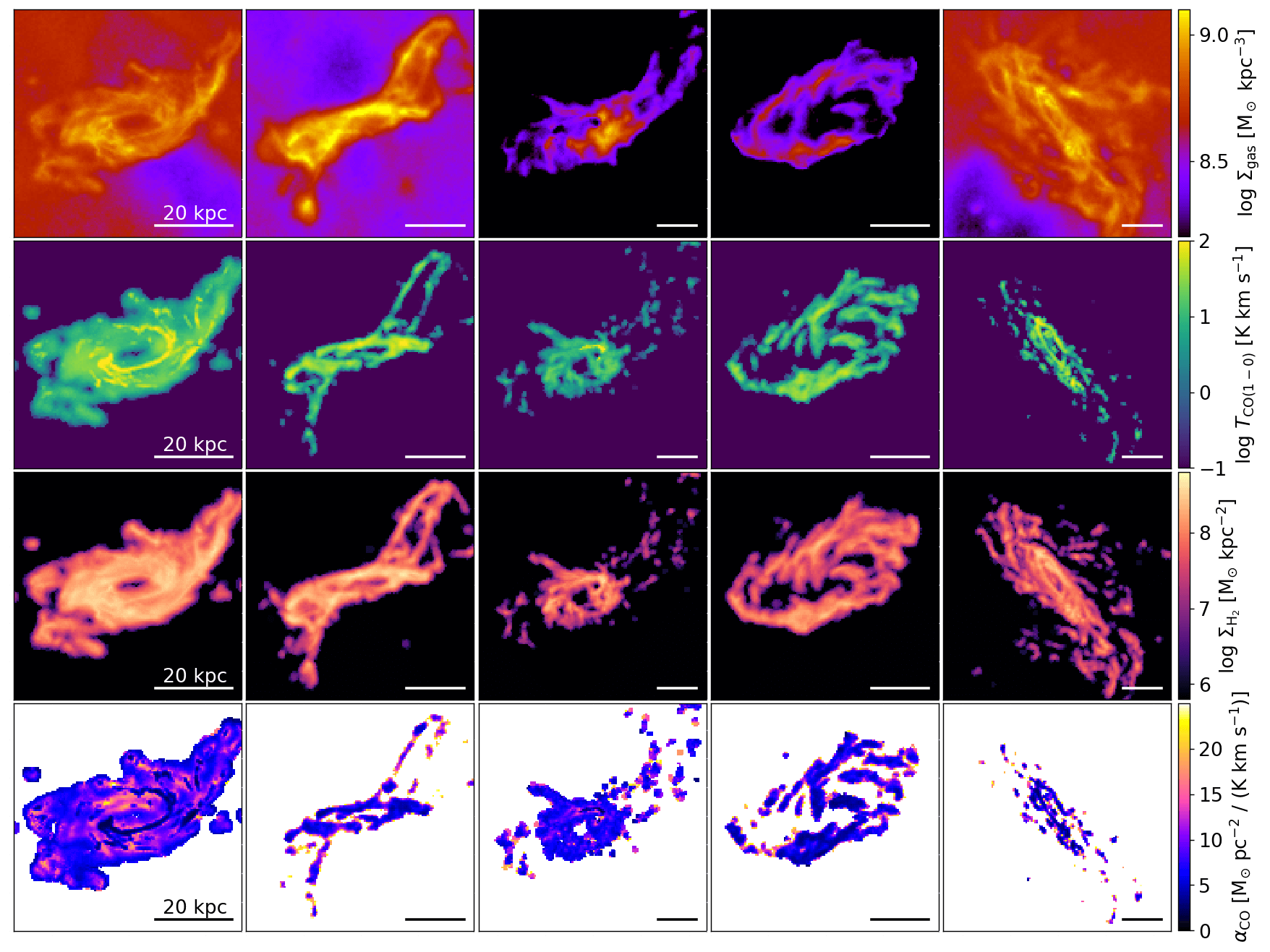}
  \caption{The large-scale distribution and the structure of galaxies. \textit{Top two panels:} stellar mass (left) and CO(1-0) luminosity (right) distributions in the whole simulation box of the TNG100-1 at $z=0$. \textit{Bottom set of panels:} the five brightest galaxies in CO(1-0) luminosity in the simulation (from left to right for the first to fifth brightest ones). From top to bottom, the panels show gas surface densities, velocity-integrated surface brightness temperatures in CO(1-0), \H2 surface densities and local $\alpha_{\rm CO}$, respectively. The galaxies are oriented in random directions. The horizontal bar on the bottom right corner in each panel indicates the physical scale of $20~{\rm kpc}$.}   
  \label{COuniverse}
\end{figure*}
The number of clouds in a gas cell is given by
 \begin{equation}
N_{\rm cloud}=\frac{V_{\rm CNM}}{V_{\rm cloud}}=\frac{m_{\rm CNM}}{m_{\rm cloud}}.
\label{n_clouds}
\end{equation}
 We allow this value to be less than $1$, i.e. a computational cell covers only a fraction of a MGC. Histograms in Fig. \ref{Ncellhist} illustrate the distribution of gas cells as functions of $N_{\rm cloud}$ in the snapshots at $z=0$, $1.0$ and $3.8$. The majority of the cells have $N_{\rm cloud}\sim2$ in all of the snapshots although the fraction of cells with $N_{\rm cloud}<1$ somewhat increases at $z=3.8$. 
By calculating a velocity-integrated line emissivity $W$ in a manner described in Section \ref{despotic}, we obtain the line intensity as $I_{\rm cloud}=\pi r_{\rm cloud}^2 W$. Then, the total intensity of a gas cell is simply
\begin{equation}
I_{\rm cell}=I_{\rm cloud} N_{\rm cloud}.
\label{cellluminosity}
\end{equation}
The total \H2 mass is $m_{\rm H_2, cell}=f_{\rm H_2}m_{\rm CNM}$, where $f_{\rm H_2}$ is mass fraction of \H2 with respect to all the components including hydrogen, helium and heavy elements. We note that introducing $N_{\rm cloud}$ in Equation (\ref{n_clouds}) enables our model to be independent of the mass-resolution of a simulation.

\section{Results}
\label{results}
\subsection{Galaxies in the local Universe}
\label{lowz}
We begin with testing our model prediction at $z=0$ by comparing with observations of local galaxies. The top two panels in Fig. \ref{COuniverse} show the large-scale distribution of stellar mass (left) and velocity-integrated CO(1-0) luminosity. The filamentary cosmic web can be seen clearly in not only the stellar distribution but also the CO(1-0) emission. The bottom set of panels show, for the five CO-brightest galaxies, the surface gas density, CO(1-0) brightness temperature $T_{\rm CO(1-0)}$, \H2 surface density, and $\alpha_{\rm CO}=\Sigma_{\rm H_2}/T_{\rm CO(1-0)}$. These CO-brightest galaxies have {\it local} values of $\alpha_{\rm CO} \sim5$--$10$. We find approximately uniform $\alpha_{\rm CO}$ in regions where CO(1-0) line is strong. Note, however, that our model does not incorporate the local variation of ISRF within a galaxy (see Section \ref{isrf}).

\subsubsection{Comparison with xCOLD GASS}
\label{xcoldgass}
\begin{figure}
  \includegraphics[bb=0 0 734 486, width=\hsize]{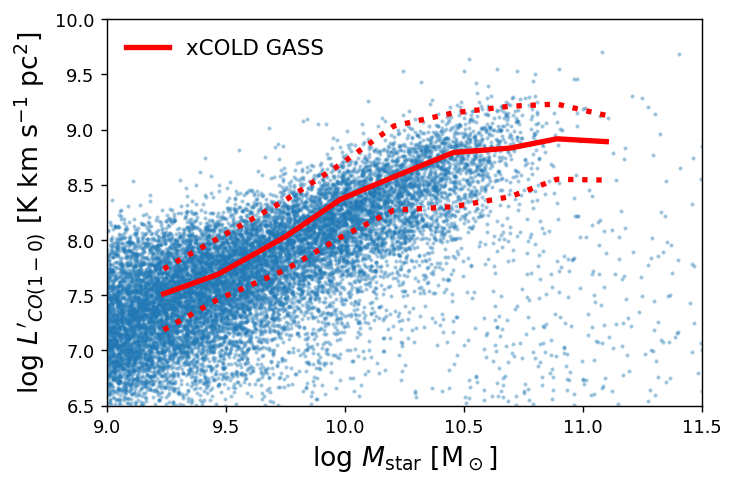}
  \caption{Velocity-integrated brightness temperatures of CO(1-0) lines of galaxies as functions of stellar mass. The blue dots indicate our results using TNG100-1 at $z=0$. The red solid and dashed lines show the median and the $\pm1\sigma$ ranges of the xCOLD GASS sample with detection of the CO(1-0) lines.}
  \label{MsLco}
\end{figure}
\begin{figure}
  \includegraphics[bb=0 0 613 588, width=\hsize]{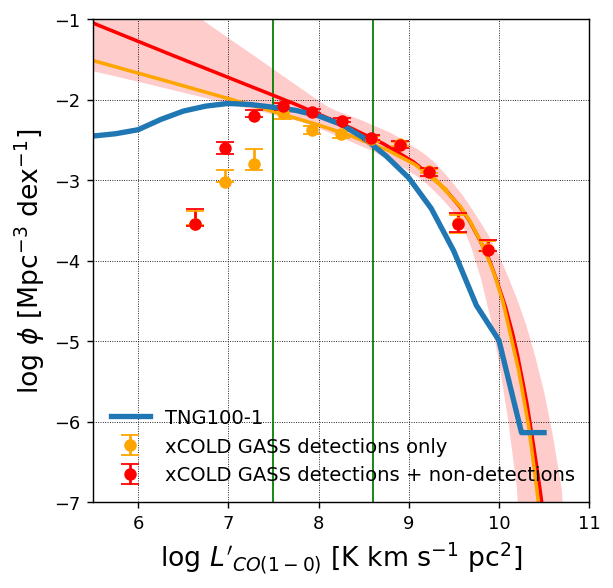}
  \caption{LFs of CO(1-0) lines at $z=0$. The blue solid line delineates our result using TNG100-1. The filled circles with error bars indicate the observations of xCOLD GASS. The yellow ones include all observed sample, but the red ones only include those with detections of CO(1-0). The yellow and red solid lines are their fittings with Schechter functions. The vertical green lines at $\log L'_{\rm CO(1-0)}=7.5$ and $8.6$ indicate the completeness limits of xCOLD GASS due to their stellar mass cut and their gas fraction integration limit, respectively. }
  \label{LcoFunction_z0}
\end{figure}
The xCOLD GASS survey \citep{xclodgass:17} has observed CO emission of local galaxies and built a large sample that is unbiased except for sampling with an equal frequency among stellar mass bins between $M_{\rm star}=10^9$ and $10^{11.5}~{\rm M_\odot}$. We compare our model prediction with their observations. Fig. \ref{MsLco} shows galaxy-integrated brightness temperatures of CO(1-0) lines as a function of stellar mass. The blue dots indicate results (galaxies) from our model, and the red lines delineate the median (solid) and $\pm1\sigma$ deviations (dotted) for the observed sample with CO(1-0) line detection. Clearly, our model reproduces the correlation of $L'_{\rm CO(1-0)}$ with $M_{\rm star}$, in agreement with the xCOLD GASS sample. The $L'_{\rm CO(1-0)}$-$M_{\rm star}$ relation is thought to correspond to the SF main sequence, i.e. correlation between galactic SFRs and $M_{\rm star}$. Galaxies with high SFRs are molecular-rich and therefore bright in CO emission.

Fig. \ref{LcoFunction_z0} compares CO-luminosity functions (LFs) between our model and xCOLD GASS. Again, the result is consistent with the observed CO-LFs, with the model underpredicting slightly in the range of $L'_{\rm CO(1-0)}\sim10^9$--$10^{10}~{\rm K~km~s^{-1}~pc^2}$. Below the completeness limit due to the stellar mass cut of xCOLD GASS at $\log L'_{\rm CO(1-0)}\lesssim7.5$, the CO-LF predicted by our model decreases and significantly deviates from the extrapolation of the Schechter functions fitted to the observations. Also in the low-luminosity range, the data points of xCOLD GASS indicate lower number density than our model. This could be attributed to the stellar mass cut of xCOLD GASS, whereas our model does not impose such a lower limit of stellar mass on our galaxy sampling. 

\begin{figure}
  \includegraphics[bb=0 0 755 594, width=\hsize]{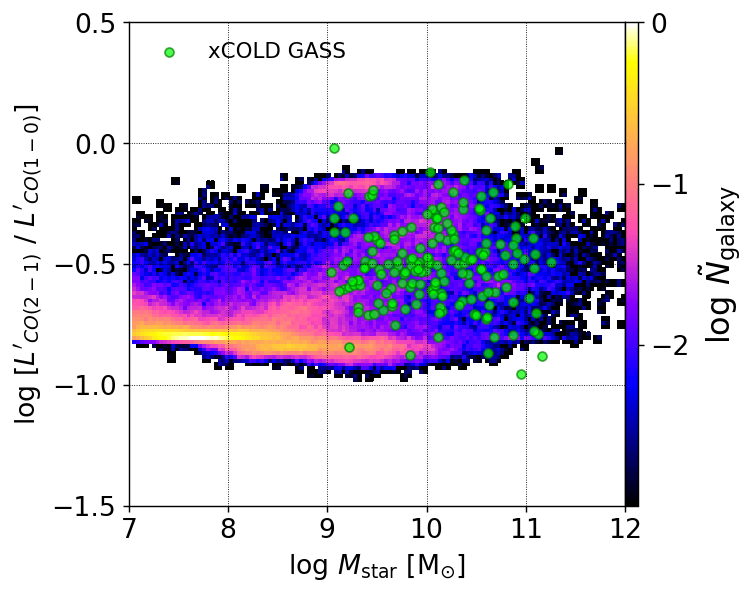}
  \caption{The line luminosity ratios $L'_{\rm CO(2-1)}/L'_{\rm CO(1-0)}$ against stellar mass. The colour indicates the number of galaxies in each bin normalised to the highest value (see the colour bar on the right). The green filled circles are the observed values of the xCOLD GASS sample.}
  \label{CO21_z0}
\end{figure}
Fig. \ref{CO21_z0} shows galaxy-integrated line ratios of $r_{21}\equiv L'_{\rm CO(2-1)}/L'_{\rm CO(1-0)}$ as a function of $M_{\rm star}$. All the simulated galaxies are located in the range between $\log r_{21}\simeq-1$ and $0$, and cover the distribution of most of the galaxies observed in xCOLD GASS in the range of $M_{\rm star}\gtrsim10^9~{\rm M_\odot}$. Interestingly, below the stellar mass limit of xCOLD GASS, i.e. in $M_{\rm star}<10^9~{\rm M_\odot}$, our result predicts that most of the low-mass galaxies have low ratios of  $\log r_{21}\simeq-0.8$. The combination of our model and the IllustirsTNG simulation can be tested by future high-sensitivity observations.

\begin{figure}
  \includegraphics[bb=0 0 710 594, width=\hsize]{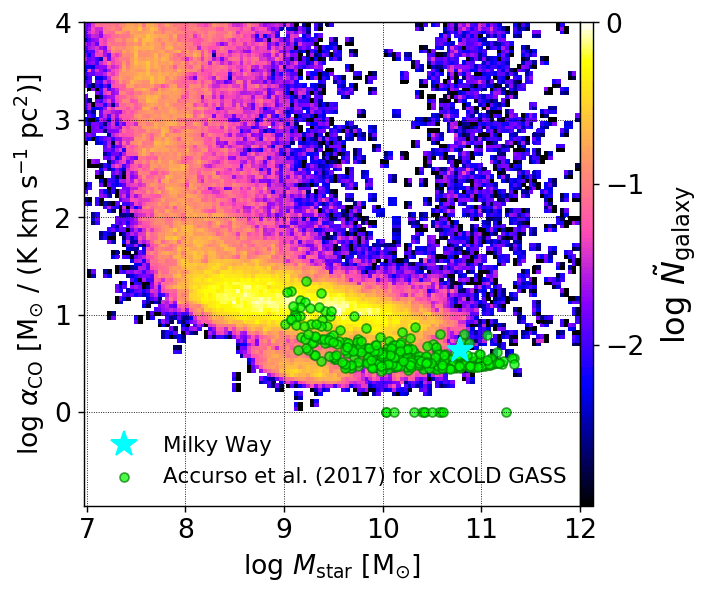}
  \caption{Distribution of galaxy-integrated $\alpha_{\rm CO}$ and stellar mass in TNG100-1. The colour code is the same as in Fig. \ref{CO21_z0}. To ensure the accuracy of $\alpha_{\rm CO}$, we here exclude the galaxies whose total masses of star-forming gas are lower than $10^7~{\rm M_\odot}$. The cyan symbol at $(\log M_{\rm star},\log\alpha_{\rm CO})=(10.8, 0.63)$ corresponds to the values measured in the inner Galactic disc. The green filled circles are $\alpha_{\rm CO}$ estimated using a model of \citet{asc:17} for the xCOLD GASS sample.}
  \label{AlphaCO_z0}
\end{figure}

Hereafter, we re-define $\alpha_{\rm CO}$ as a galaxy-integrated value: $\alpha_{\rm CO} \equiv M_{\rm H_2}/L'_{\rm CO(1-0)}$. We note that $\alpha_{\rm CO}$ in the bottom panels of Fig. \ref{COuniverse} is defined as the ratio of surface \H2 mass density to CO(1-0) brightness temperature measured locally. Fig. \ref{AlphaCO_z0} shows the distribution of $\alpha_{\rm CO}$ and galactic stellar masses. We find $\alpha_{\rm CO}$ is approximately constant at $\log\alpha_{\rm CO}\simeq1$ in the range of $M_{\rm star}\sim10^8$--$10^{10.5}~{\rm M_{\odot}}$. However, $\alpha_{\rm CO}$ increases below $M_{\rm star}\sim10^8~{\rm M_{\odot}}$, and there is a population of massive galaxies that have high $\alpha_{\rm CO}$ with $M_{\rm star}\gtrsim10^{11}~{\rm M_{\odot}}$. These low- and high-mass galaxies with such high $\alpha_{\rm CO}$ correspond to dwarf and massive elliptical galaxies, and their high $\alpha_{\rm CO}$ are because of their low metallicities and/or diffuse gas distribution (see Section \ref{narayanan}). In Fig. \ref{AlphaCO_z0}, the cyan star-shaped symbol indicates $\alpha_{\rm CO}=4.3$ measured in the Galactic inner disc \citep{bwl:13} with the Galactic stellar mass of $M_{\rm star}=6.08\times10^{10}~{\rm M_{\odot}}$ \citep{ln:15}. The green filled  circles correspond to the xCOLD GASS sample where $\alpha_{\rm CO}$ are not observationally determined but estimated using the model of \citet{asc:17}.\footnote{The model of \citet{asc:17} gives $\alpha_{\rm CO}$ as a function of metallicity and offset from the SF main sequence.} These values of $\alpha_{\rm CO}$ appear to be somewhat lower than the averaged values in our model although these are within the range covered by our prediction.

\begin{figure}
  \includegraphics[bb=0 0 716 486, width=\hsize]{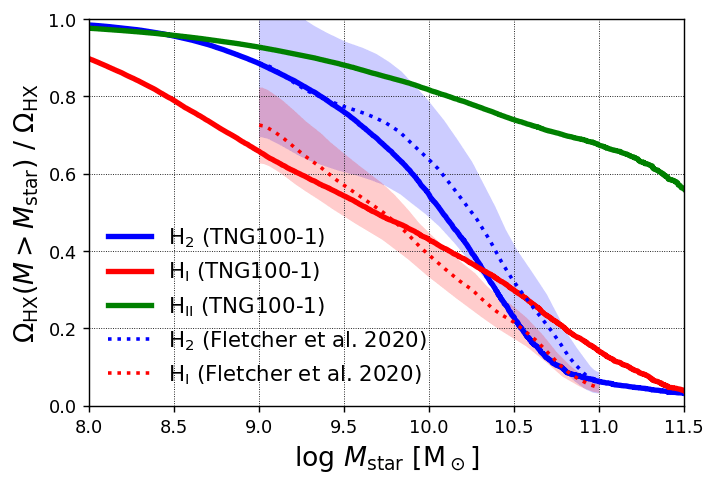}
  \caption{The fractions of \H2 (blue), H$_{\rm I}$ (red) and H$_{\rm II}$ (green) that reside in galaxies above a given $M_{\rm star}$. The solid lines indicate our results. The dotted lines are the observational results of \citet{fss:20} derived from the xCOLD GASS (blue dotted) and xGASS (red dotted) surveys, and the shaded regions show their observational $1\sigma$ errors.}
  \label{Omega_z0}
\end{figure}
From our model, we can compute not only mass of \H2 but also those of H$_{\rm I}$ and H$_{\rm II}$. We assume that the star-forming gas is neutral. In such a gas cell, WNM is assumed to be fully atomic, and abundances of \H2 and H$_{\rm I}$ in CNM are computed with {\sc DESPOTIC}.\footnote{Although {\sc DESPOTIC} also computes H$_{\rm II}$ abundance in a MGC, an amount of H$_{\rm II}$ is generally negligible.} We consider the diffuse gas with $\rho_{\rm cell}<n_{\rm H,SF}$ to form no molecules, and its ionised fraction is computed in the IllustrisTNG simulation. Note again that we do not take into account inter-galactic gas that is not bound to any galaxies. Using these quantities, we estimate the cosmic densities of hydrogen in molecular, atomic and isonised states to be $\Omega_{\rm H_2}=6.44\times10^{-5}h^{-1}$, $\Omega_{\rm H_I}=4.13\times10^{-4}h^{-1}$ and $\Omega_{\rm H_{II}}=4.79\times10^{-3}h^{-1}$, respectively. In observations, \citet{fss:20} estimates $\Omega_{\rm H_2}=(5.34\pm0.47)\times10^{-5}h^{-1}$ from xCOLD GASS and $\Omega_{\rm H_I}=(2.35^{+2.17}_{-0.67})\times10^{-4}h^{-1}$ from the xGASS survey \citep{csj:18}. \citet{jhg:18} determine $\Omega_{\rm H_I}=(3.8\pm0.7)\times10^{-4}h^{-1}$ from the ALFALFA survey \citep{ghk:05}. The values of $\Omega_{\rm H_2}$ and $\Omega_{\rm H_I}$ in our model are thus consistent with these observational measurements although $\Omega_{\rm H_2}$ is slightly above the error range of the xCOLD GASS observations. Fig. \ref{Omega_z0} shows fractions of the cosmic hydrogen densities cumulated from galaxies with high $M_{\rm star}$. The fractions of $\Omega_{\rm H_2}$ in our model is consistent with the observations of \citet{fss:20} within the error ranges although those of $\Omega_{\rm H_I}$ appear to be somewhat higher than the observations in $M_{\rm star}\gtrsim10^{10.5}~{\rm M_\odot}$. As \citet{fss:20} mention, our results show that nearly ninety per cent of \H2 in the Universe resides in galaxies with $M_{\rm star}>10^{9}~{\rm M_\odot}$, and dwarf galaxies do not host a significant amount of molecules. However, it is worth reminding of the fact that the observed \H2 mass is estimated from CO luminosity via $\alpha_{\rm CO}$ of \citet{asc:17}; on the other hand, our model directly computes \H2 mass. It is known that dwarf galaxies generally have low SF efficiencies leading to low stellar mass to halo mass ratios \citep[e.g.][]{bwc:13a,bwc:13b,bwh:19}, and it is often attributed to intense gas outflows by SNe due to their shallow potentials. From our result in Fig. \ref{Omega_z0}, we argue that the low \H2 abundances in the low-mass galaxies could be another cause of the low SF efficiencies of dwarfs. 

\begin{figure}
  \includegraphics[bb=0 0 700 486, width=\hsize]{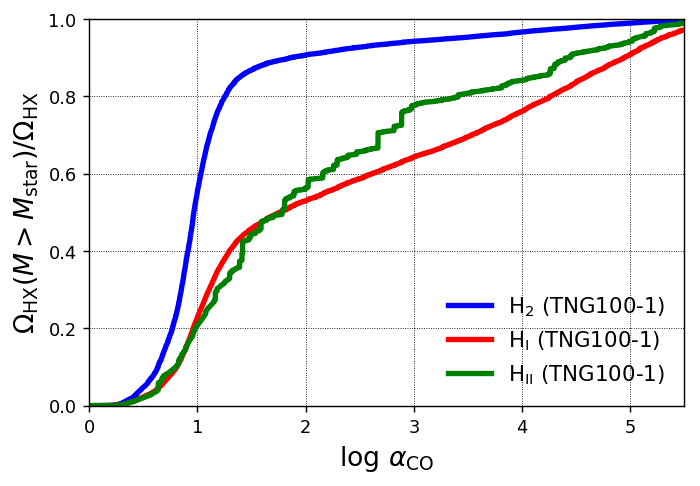}
  \caption{Same as Fig. \ref{Omega_z0} but for the fractions of hydrogen that resides in galaxies below a given $\alpha_{\rm CO}$.}
  \label{COdark_z0}
\end{figure}
Fig. \ref{COdark_z0} shows the fractions of the cosmic hydrogen densities summed over galaxies with low $\alpha_{\rm CO}$ in our model. Nearly ninety per cent of \H2 gas in the Universe resides in galaxies with $\log\alpha_{\rm CO}\simeq0.5$--$1.5$, and only ten per cent of \H2 is formed in `CO-dark' galaxies with $\log\alpha_{\rm CO}\gtrsim1.5$. Although such CO-dark molecular clouds in these galaxies would be missed in observations, their total amount is expected to be insignificant with respect to the total molecular mass in the Universe. However, the CO-dark galaxies host nearly half the total amounts of H$_{\rm I}$ and H$_{\rm II}$. 

As we show above, our model adopted to IllustrisTNG with our fiducial parameter settings can thus reproduce well the galaxy-integrated properties reported in the previous observational studies. We discuss parameter-dependence of our model in Section \ref{discussz0}.

\subsubsection{Other models}
\label{narayanan}
\begin{figure}
  \includegraphics[bb=0 0 820 691, width=\hsize]{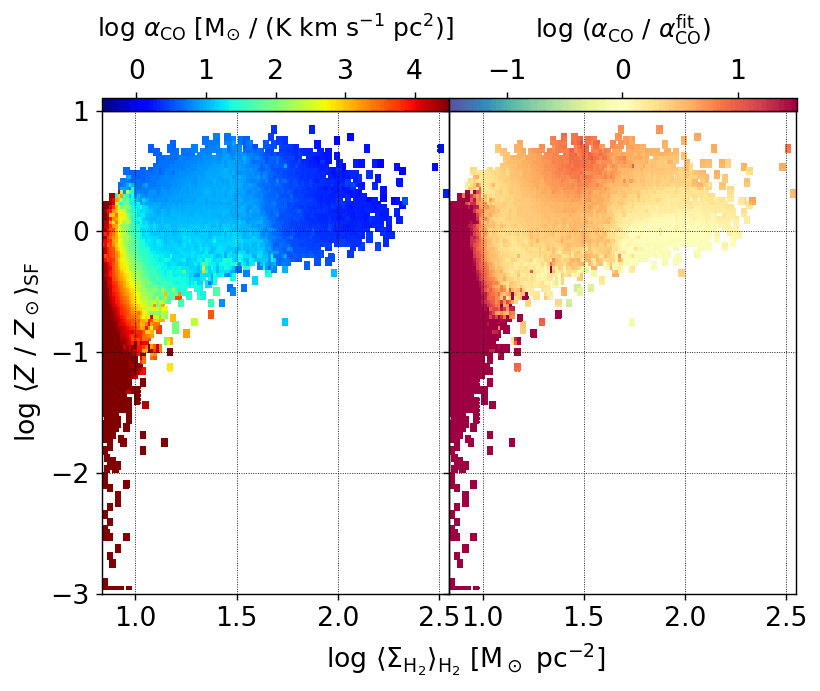}
  \includegraphics[bb=0 0 835 691, width=\hsize]{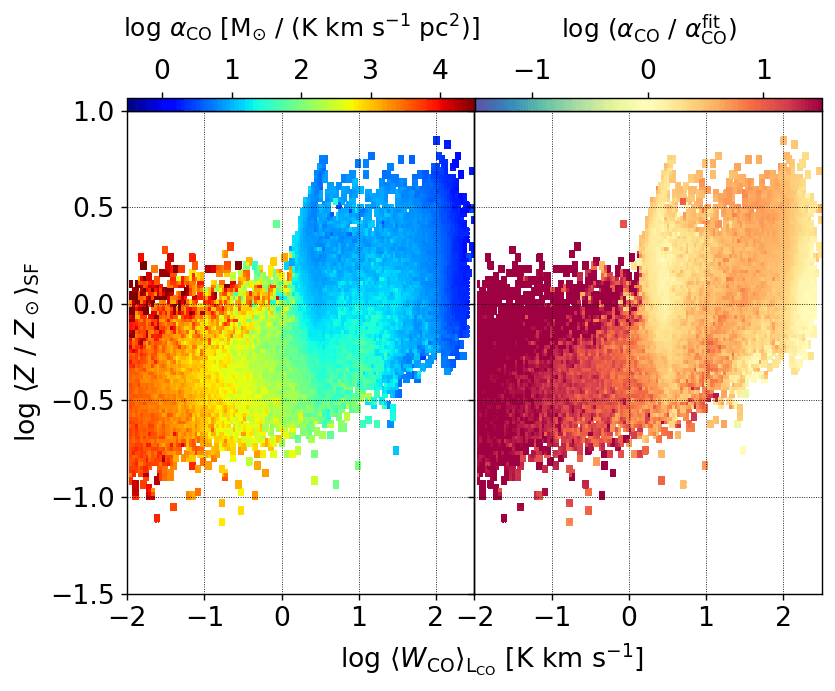}
  \caption{\textit{Left panels:} the averaged values of $\log\alpha_{\rm CO}$ in our model. \textit{Right panels:} the differences of our results from the fitting functions of Equations (\ref{Narayanan1} and \ref{Narayanan2}); if $\log(\alpha_{\rm CO}/\alpha_{\rm CO}^{\rm fit})>0$, our model predicts higher $\alpha_{\rm CO}$ than the fitting functions. The ordinates $\langle Z/Z_\odot\rangle_{\rm SF}$ indicate mass-weighted metallicities among star-forming gas cells within the galaxy. In the top panels, the abscissa $\langle\Sigma_{\rm H_2}\rangle_{\rm H_2}$ indicates average of \H2 column densities weighted by molecular mass over all gas clouds in the galaxy. In the bottom panels, the abscissa $\langle W_{\rm CO}\rangle_{\rm L_{CO}}$ indicates average of luminosity-weighted CO emissivities over all clouds in the galaxy. We here exclude galaxies whose masses of star-forming gas are lower than $10^7~{\rm M_\odot}$.}
  \label{aCORhoMet_z0}
\end{figure}
We find that the galaxy-integrated $\alpha_{\rm CO}$ strongly correlates with the averaged metallicities $Z$ of star-forming gas and their column densities $\Sigma_{\rm cloud}$, whereas the correlations with the other properties such as ISRF $\chi_{\rm cloud}$ and gas fractions are less clear. This finding is consistent with the result of \citet{nko:12}, in which they have proposed a fitting function obtained from their isolated and merger simulations:
\begin{equation}
\alpha_{\rm CO}^{\rm fit} = \frac{20.6}{\langle Z/Z_\odot\rangle\langle\Sigma_{\rm H_2}\rangle_{\rm H_2}^{0.5}},
\label{Narayanan1}
\end{equation}
where $\langle Z/Z_\odot\rangle$ is mass-weighted mean of gaseous metallicity, and $\langle\Sigma_{\rm H_2}\rangle_{\rm H_2}$ is average of \H2 column densities weighted by \H2 mass over all gas clouds\footnote{The definition is $\langle\Sigma_{\rm H_2}\rangle_{\rm H_2}\equiv\sum_im_{{\rm H_2},i}\Sigma_{{\rm H_2},i}/\sum_im_{{\rm H_2},i}$, where $m_{{\rm H_2},i}$ is \H2 mass in $i$-th gas cell, and $\Sigma_{{\rm H_2},i}$ is \H2 column density of a gas cloud in the cell: $\Sigma_{{\rm H_2}}=\Sigma_{\rm cloud}f_{\rm H_2}$.} in the units of ${\rm M_\odot~pc^{-2}}$. The top panels of Fig. \ref{aCORhoMet_z0} show distribution of the ensemble averages of $\log\alpha_{\rm CO}$ in our model and comparison with the fitting function of equation (\ref{Narayanan1}). In the top left panel, the mean $\alpha_{\rm CO}$ increases with decreasing $Z$ and $\Sigma_{\rm H_2}$. Especially, most of the metal-poor galaxies with $\log\langle Z/Z_\odot\rangle\lesssim-0.5$ have quite high $\alpha_{\rm CO}$, and galaxies with $\langle\Sigma_{\rm H_2}\rangle_{\rm H_2}\lesssim10~{\rm M_\odot~pc^{-2}}$ have high $\alpha_{\rm CO}$ irrespective of their metallicities. In the top right panel, we show logarithmic differences of $\alpha_{\rm CO}$ between our model and equation (\ref{Narayanan1}). The values of $\alpha_{\rm CO}$ are consistent between our model and the fitting function for the galaxies with $\langle\Sigma_{\rm H_2}\rangle_{\rm H_2}\gtrsim10~{\rm M_\odot~pc^{-2}}$ within a factor of unity. The fitting function, however, predicts significantly lower $\alpha_{\rm CO}$ than our model for the galaxies with $\langle\Sigma_{\rm H_2}\rangle_{\rm H_2}\lesssim10~{\rm M_\odot~pc^{-2}}$. This may be because the fitting function is derived from a number of their isolated and merger simulations for relatively massive galaxies with $M_{\rm baryon}\sim10^{11}~{\rm M_\odot}$.

The fitting function of Equation (\ref{Narayanan1}) can be converted to an `observable form' \citep{nko:12},
\begin{equation}
\alpha_{\rm CO}^{\rm fit} = \frac{10.7}{\langle Z/Z_\odot\rangle^{0.65}\langle W_{\rm CO}\rangle_{\rm L_{CO}}^{0.32}},
\label{Narayanan2}
\end{equation}
where $\langle W_{\rm CO}\rangle_{\rm L_{CO}}$ is the luminosity-weighted mean of CO(1-0) emissivities over all clouds in a galaxy in units of ${\rm K~km~s^{-1}}$. The bottom panels of Fig. \ref{aCORhoMet_z0} show the same as the top ones but with $\langle W_{\rm CO}\rangle_{\rm L_{CO}}$ for the abscissas. The values of $\alpha_{\rm CO}$ increase with decreasing $\langle W_{\rm CO}\rangle_{\rm L_{CO}}$ in our model, and equation (\ref{Narayanan2}) gives significantly lower $\alpha_{\rm CO}$ for galaxies with low $\langle W_{\rm CO}\rangle_{\rm L_{CO}}\lesssim1~{\rm K~km~s^{-1}}$. Thus, the fitting functions derived by \citet{nko:12} appear to be accurate and useful for CO-bright galaxies that have dense molecular clouds with $\langle W_{\rm CO}\rangle_{\rm L_{CO}}\gtrsim1~{\rm K~km~s^{-1}}$ and $\langle\Sigma_{\rm H_2}\rangle_{\rm H_2}\gtrsim10~{\rm M_\odot~pc^{-2}}$. 

Our model employs a simple approximation for calculating ISRF $\chi_{\rm cloud}$ and dust opacity $\tau$ for gas clouds (Section \ref{isrf}). \citet{nko:12} use dust radiation transfer calculations \citep{nko:11} for spatially resolved distribution of radiation sources such as stars and active galactic nuclei with spectrum energy distribution models. It is noteworthy that, despite the simplicity, our model reproduces their fitting results of $\alpha_{\rm CO}^{\rm fit}$ for a large number of galaxies in cosmological simulations except for the diffuse galaxies with $\langle\Sigma_{\rm H_2}\rangle_{\rm H_2}\lesssim10~{\rm M_\odot~pc^{-2}}$.  

\subsection{Results at high redshifts}
\label{highz}
\begin{figure*}
  \includegraphics[bb=0 0 1494 387, width=\hsize]{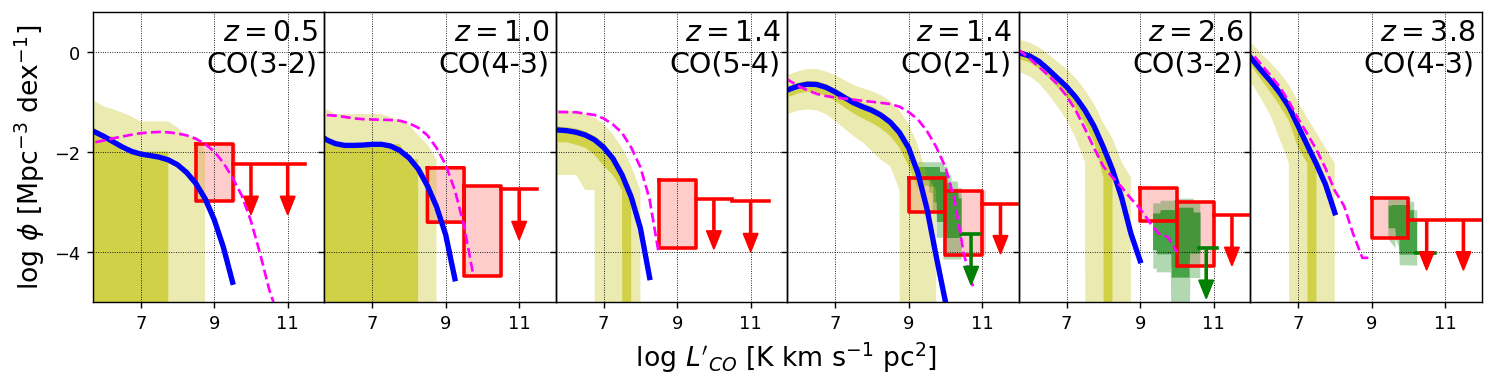}
  \caption{Luminosity functions of CO[$J$-($J-1$)] transition lines, where $J$ is consistent with that of the ASPECS surveys in each panel. The blue solid lines indicate our results including all galaxies in the whole simulation box of TNG100-1. The shaded regions with thick and thin yellow are $1\sigma$ and $2\sigma$ confidence intervals of the cosmic variance computed in the volumes whose sizes are the same as the observing regions of ASPECS. The magenta dashed lines are the same as the blue solid lines but assuming the cloud sizes to be ten times larger: $r_{\rm cloud}=5\lambda_{\rm J}$ (see Section \ref{discusshighz}). The red and green boxes indicate the observational determinations of the ASPECS pilot and LP \citep[see][]{aspecs_pilot,aspecs_LP}, respectively. The vertical sizes of the red boxes show the Poisson errors of the pilot observation. The thick and thin green regions correspond to $1\sigma$ and $2\sigma$ confidence intervals of the LP observations. The horizontal bars with downwards arrows indicate upper limits of the observations.}   
  \label{LcoFuncHighJ}
\end{figure*} 
We post-process snapshots of TNG100-1 at high redshifts. From the agreement of our results with the xCOLD GASS survey in Section \ref{xcoldgass}, we can possibly apply our model to high-redshift galaxies without calibrating the parameters or altering our model. This may, however, be a naive expectation. For example, dust properties such as typical size $a$ and dust-to-metal fraction $f_{\rm dust}$ (Sections \ref{isrf} and \ref{despotic}) can evolve with redshift. The inaccuracy of our modelling for ISRF $\chi_{\rm cloud}$ can be amplified since the ISRF can be more intense due to the higher SFRs and lower dust abundances (lower metallicities) of high-redshift galaxies. In addition, galactic morphologies are more irregular and complex due to intense gas accretion and frequent mergers at higher redshifts. Our modelling such as assuming $r_{\rm cloud}=\lambda_{\rm J}/2$ may no longer be accurate for such dense and irregular galaxies.

We adopt the fiducial model described in Section \ref{cloudmodel} to outputs of TNG100-1 from $z=0.5$ to $3.8$ without changing the parameters. In what follows, we examine (in)consistency with CO line observations at high redshifts. We note that the available observations are still limited to a small number of galaxies. Future large, systematic surveys will allow us
to perform substantially better and rigorous comparison.

\subsubsection{Comparison with ASPECS}
\label{aspecs}
We compare our results from the ASPECS surveys. \citet{aspecs_pilot} provide the results of their pilot observation at wavelengths of $1$ and $3~{\rm mm}$ (band 3 and 6) for galaxies at the mean redshifts $\langle z\rangle=0.5$, $1.0$, $1.4$, $2.6$ and $3.8$. \citet{aspecs_LP} present data from large-programme (LP) observations at $3~{\rm mm}$ for galaxies at $\langle z \rangle=0.3$, $1.4$, $2.6$ and $3.8$, which covers a wider area than the pilot survey. Redshifted CO(1-0) emission (rest-frame $2.6~{\rm mm}$ line) is not covered by currently available receiver bands of ALMA. Hence, the high-redshift CO observations are aimed at detecting higher-$J$ lines, CO[$J$-($J-1$)], that have shorter rest-frame wavelengths. The ASPECS surveys observe the emission of $J=3$, $4$, $2$, $3$ and $4$ at $z=0.5$, $1.0$, $1.4$, $2.6$ and $3.8$, respectively; CO(5-4) is also observed at $z=1.4$ in the pilot observation. 

In our model, high-$J$ lines are directly calculated with {\sc DESPOTIC} without introducing any conversion factors. The effective survey volume of ASPECS is much smaller than the simulation box of TNG100-1. We are thus able to estimate the cosmic variance by sampling cubic regions with side lengths of $4.59$, $5.18$, $8.30$, $6.82$ and $5.48~{\rm Mpc}$ (physical) at $z=0.5$, $1.0$, $1.4$, $2.6$ and $3.8$, respectively. These sampling volumes are consistent with those of ASPECS pilot observation for $z=0.5$, $1.0$ and APECS LP for $z=1.4$, $2.6$ and $3.8$. A total of 16384 cubic regions are randomly selected in the simulation box. 

\begin{figure*}
  \includegraphics[bb=0 0 1494 388, width=\hsize]{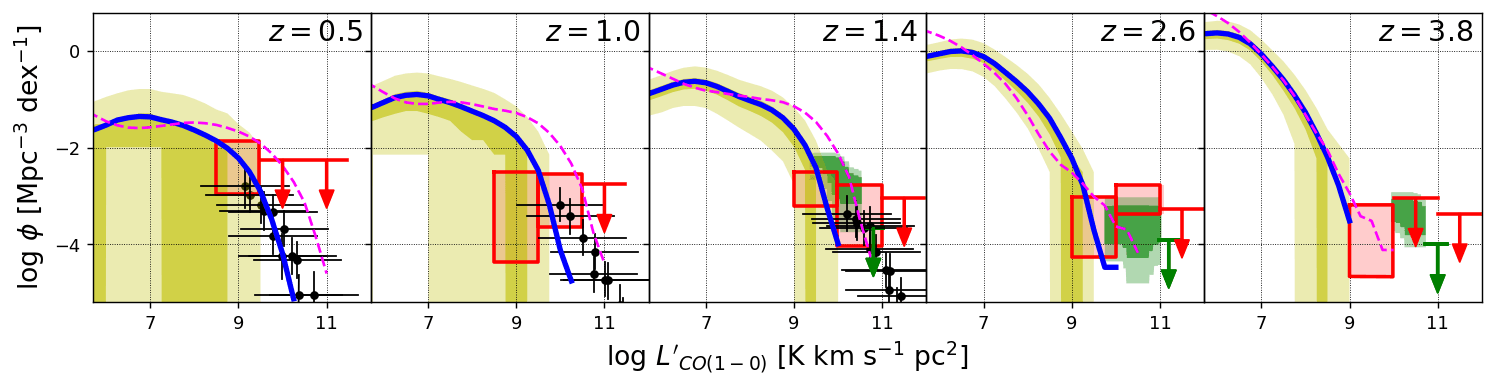}
  \caption{Same as Fig. \ref{LcoFuncHighJ} but for CO(1-0) emission. In the three panels from left, the black filled circles with error bars indicate the observational results of \citet{oil:20} at $z=0.5$, $1.1$ and $1.5$, respectively.}   
  \label{LcoFuncCO10}
\end{figure*}
\begin{figure*}
  \includegraphics[bb=0 0 1494 384, width=\hsize]{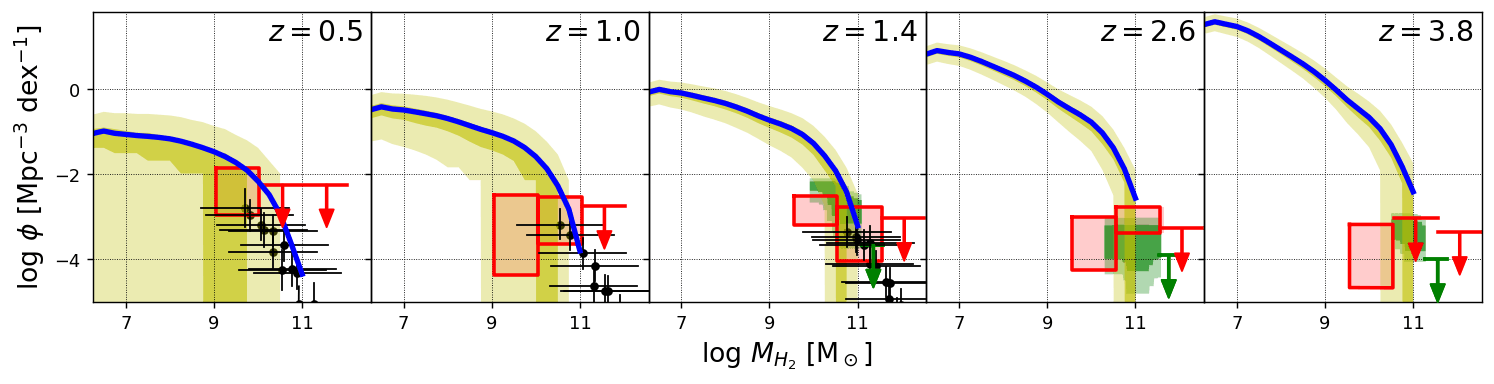}
  \caption{Same as Figs. \ref{LcoFuncHighJ} and \ref{LcoFuncCO10} but for \H2 masses. The observations plotted here are the same as those in Fig. \ref{LcoFuncCO10} but shifted horizontally by $\alpha_{\rm CO}=3.6$. The results with $r_{\rm cloud}=5\lambda_{\rm J}$ are not shown since these are hardly different from the fiducial case.}   
  \label{H2Func}
\end{figure*}
Fig. \ref{LcoFuncHighJ} compares high-$J$ CO-LFs. The blue solid lines indicate our results including all the galaxies in the simulation. Our model appears to underpredict the CO-LFs for CO(5-4) at $z=1.4$, CO(3-2) at $z=2.6$ and CO(3-2) at $z=3.8$. The model CO-LFs are outside the ranges of uncertainties of the ASPECS pilot and LP data.

For high-redshift galaxies, CO(1-0) line luminosity is {\it estimated} from the higher-$J$ lines using the line ratios $r_{J1}$ for an assumed line energy distribution: $L'_{\rm CO[J-(J-1)]}=r_{J1}L'_{\rm CO(1-0)}$. The ASPECS surveys use $r_{J1}$ estimated by \citet{ddl:15}: $r_{J1}=0.76$, $0.42$, $0.31$ and $0.23$ for $J=2$, $3$, $4$ and $5$, respectively. Our model provides CO(1-0) line luminosities directly. Fig. \ref{LcoFuncCO10} compares our model CO(1-0) LFs with those converted from the ASPECS data. We also plot the observational results of \citet{oil:20} which include brighter galaxies than ASPECS.\footnote{\citet{oil:20} estimate the CO-LFs for their compilation of various observations using correlations between the luminosities of radio continuum, infrared and CO emission.} Our model matches the observed CO-LFs within the error ranges at $z\lesssim2.0$. At $z=2.6$ and $3.8$, the modelled galaxies appear to lack galaxies brighter than $L'_{\rm CO(1-0)}=10^9~{\rm K~km~s^{-1}~pc^2}$, which is clearly inconsistent with the ASPECS surveys.

The ASPECS observations assume $\alpha_{\rm CO}=3.6~{\rm M_\odot~(K~km~s^{-1}~pc^2)^{-1}}$ for all observed galaxies.\footnote{\citet{oil:20} also assume $\alpha_{\rm CO}=3.6$.} Accordingly, their galaxy-integrated \H2 mass is 
estimated as 
\begin{equation}
M_{\rm H_2}=\frac{\alpha_{\rm CO}}{r_{J1}}L'_{\rm CO[J-(J-1)]}.
\label{MH2estimate}
\end{equation}
Note the estimate is subject to uncertainties of the two conversion factors: $\alpha_{\rm CO}$ and $r_{J1}$, whereas our model can calculate $M_{\rm H_2}$ independently from CO line luminosity. Fig. \ref{H2Func} compares the model and observed (estimated) \H2 mass functions.  The model  agrees reasonably well with ASPECS and \citet{oil:20} at $z=0.5$, $1.0$ and $1.4$. However, our model tends to overpredict $M_{\rm H_2}$ at $z>2$. This trend is actually {\it opposite} to what is found for CO LFs. Overall, at high redshift, our model predicts lower $L'_{\rm CO}$ and larger $M_{\rm H_2}$ than ASPECS, and the discrepancy appears to be larger at higher redshift. This indicates that the average $\alpha_{\rm CO}$
 in our model increases with redshift. In Section \ref{discusshighz}, we discuss possible causes of the variation of $\alpha_{\rm CO}$ and further investigate the inconsistency with the observations.

\section{Discussion}
\label{discuss}
\subsection{Parameter calibration at $z=0$}
\label{discussz0}
The results presented in Section \ref{xcoldgass} show that our model reproduces well the galaxy-integrated luminosities of CO(1-0) and the line ratios to CO(2-1) of the xCOLD GASS sample. From the consistency of the molecular mass distribution shown in Fig. \ref{Omega_z0}, we expect that $\alpha_{\rm CO}$ in our model is close to those estimated for the observed galaxies (see also Fig. \ref{AlphaCO_z0}). 

Since our post-processing method employs several controlling parameters described in Section \ref{simandmethod}, we discuss the dependence of our results on the parameters. In Section \ref{cloudmodel}, we assume the density ratio based on the two-phase ISM model with $\phi\equiv\rho_{\rm CNM}/\rho_{\rm WNM}=100$, which yields the density enhancement factor of $1<f_{\rm M}/f_{\rm V}\lesssim20$. If we set $f_{\rm M}=f_{\rm V}=1$, by effectively deviating from the ISM model, our results of $L'_{\rm CO}$ and $M_{\rm H_2}$ do not systematically change. Only the scatter becomes large for low-mass galaxies. This is because the most important parameter for molecular abundance in a cloud is $\Sigma_{\rm cloud}$ rather than $\rho_{\rm CNM}$ \citep[e.g.][]{lnd:18}. We approximate a cloud size to be $r_{\rm cloud}=\lambda_{\rm J}/2$, and $\lambda_{\rm J}\propto\rho_{\rm CNM}^{-1}$ under the pressure equilibrium between the CNM and WNM. Therefore, $\Sigma_{\rm cloud}\propto\rho_{\rm CNM} r_{\rm cloud}$ is independent of the density enhancement factor, and the resulting $L'_{\rm CO}$ and $M_{\rm H_2}$ do not change significantly. In fact, we find that increasing $\Sigma_{\rm cloud}$ can make $L'_{\rm CO}$ considerably larger (see Section \ref{discusshighz}).

We employ a simple model for the unattenuated ISRF $\chi_{\rm int}$ that is proportional to the total SFR in a galaxy (equation \ref{intISRF}). To test the effect of the ISRF intensity, we alter $\chi_{\rm int}$ to be proportional to SFR$\times(r_{\rm s}/r_{\rm SFR})^{2}$ where $r_{\rm s}=3$ and $10~{\rm kpc}$. In these test models, the ISRF is stronger in a more compact galaxy for a given SFR. We find that this alteration only decreases $L'_{\rm CO}$ and $M_{\rm H_2}$ in low-mass galaxies, and that massive galaxies are little affected. We also note that the cosmic background radiation $\chi_{\rm ext}$ is generally weaker than $\chi_{\rm int}$, and thus does not significantly impact our results. 

The dust optical depth $\tau$ is another important parameter. In equation (\ref{dusttau}), the optical depth $\tau$ is proportional to an uncertain factor $f_{\rm dust}/(as)$ that involves dust-to-metal fraction, grain size and density. Decreasing $f_{\rm dust}/(as)$ can effectively lower $L'_{\rm CO}$ because of stronger ISFR. If we assume a lower $f_{\rm dust}/(as)$ than the fiducial value, our CO-LF shown in Fig. \ref{LcoFunction_z0} becomes inconsistent with the observations. In the extreme case of $\tau=\infty$, our results at $z=0$ are still consistent with the observations within the error range (although see Section \ref{discusshighz}). 

DESPOTIC can treat a different calculation mode by, for example, assuming a spherical cloud as radially stratified multiple shells with different optical depths. The multi-zone computation is thought to be more accurate than the single-zoned model we use in this study. \citet{lnd:18} perform convergence tests using DESPOTIC and conclude that the eight-zoned model is sufficient to produce a converged result. We have repeated the same post-processing computations with the eight-zoned model, and have found that the results are essentially unchanged statistically from the fiducial case.

\subsection{Evolution of ISM properties}
\label{discusshighz}
The apparent discrepancy between our model and the ASPECS observations at high redshift (Section \ref{aspecs}) may suggest evolution of ISM structure in galaxies. At $z\gtrsim1.5$--$2$, our model predicts lower $L'_{\rm CO(1-0)}$ and  higher $M_{\rm H_2}$ than the observational estimates. Compared with ASPECS, the luminosities of high-$J$ lines $L'_{\rm CO[J-(J-1)]}$ are systematically lower at high redshift. The trend of producing large \H2 masses appears to be opposite to the result of \citet{pps:19}, who find significantly lower $M_{\rm H_2}$ than the ASPECS at high redshift, although their model reproduces the \H2 mass at low-redshift observations. This may suggest ISM structure evolution over redshift.

We identify each galaxy as a gravitational bound object, whereas radio observations capture a galaxy's CO emission within a beam size. This procedure may affect the total amount of $L'_{\rm CO(1-0)}$ and $M_{\rm H_2}$. Using the IllustrisTNG simulation, \citet{pps:19} examine the systematic `aperture bias'. The resulting \H2 mass functions are not significantly different at high redshift between the two cases of using the SUBFIND grouping and the $3.5~{\rm arcsec}$ aperture assuming the ASPECS observations.

We also find that assuming a large $\tau$ in our model does not significantly affect our CO-LFs and can only mildly reduce the difference. Even in the case of $\tau=\infty$, our model predicts lower $L'_{\rm CO}$ than the observations at high redshifts. The \H2 mass functions are hardly affected.  

It has turned out that the cloud radius is the most important quantity that affects the CO line emission strengths. In Figs. \ref{LcoFuncHighJ} and \ref{LcoFuncCO10}, we plot CO-LFs with assuming ten times larger cloud radii as $r_{\rm cloud}=5\lambda_{\rm J}$ (the magenta dashed lines). These test results are in better agreement with the ASPECS observations. However, $L'_{\rm CO(1-0)}$ at $z=0.5$ and $1.0$ are larger than the observations of \citet{oil:20} if we adopt $r_{\rm cloud}=5\lambda_{\rm J}$. Overall, we find that assuming enlarged $r_{\rm cloud}$ overpredicts the CO-LF at low redshift. Enlarging $r_{\rm cloud}$ hardly changes the \H2 mass functions. In determining $\lambda_{\rm J}$ (equation \ref{jeans}), our model does not take into account contributions by turbulent or magnetic pressure.
\footnote{If the computation of $\lambda_{\rm J}$ includes the turbulent and magnetic pressure, $\gamma P_{\rm CNM}$ in equation (\ref{jeans}) is replaced with $\gamma P_{\rm CNM} + \rho_{\rm CNM}(v_{\rm turb}^2 + v_{\rm A}^2)$, where $v_{\rm turb}$ and $v_{\rm A}$ are turbulent and Alfv\'{e}n velocities of the CNM \citep[e.g.][]{fk:12}.} 
Although the turbulent velocities in local galaxies are typically lower than the ISM sound velocity, it is known that star-forming galaxies at high redshift are highly turbulent with velocity dispersions of $\sim100~{\rm km~s^{-1}}$ \citep[e.g.][]{fgb:09}. \citet{zfy:17} show illustrative comparison of velocity dispersions between star-forming galaxies observed at low and high redshifts. Ignoring the turbulent pressure can thus underestimate $\lambda_{\rm J}$ for high-redshift galaxies. Intriguingly, large turbulent velocities are thought to produce large, massive gas clouds \citep[e.g.][]{dsc:09,drc:19}. A possible prescription to match our results to the observations may be to introduce a fudge factor to increase $r_{\rm cloud}$ depending on redshift. 

It may be possible that state-of-the-art cosmological hydrodynamics simulations are still unable to reproduce gas properties at high redshifts. The IllustrisTNG simulation reproduce a variety of statistical properties of observed galaxies such as stellar mass function, mass-metallicity relation and redshift-evolution of the cosmic SFR density \citep[e.g.][]{vgs:13,gvs:14,nps:18,npsr:18,psn:18}. However, it has not been fully examined whether the properties of the local ISM are compatible with real galaxies. \citet{iy:19} have demonstrated that clumpiness of high-redshift disc galaxies can strongly depend on the ISM model (EOS of star-forming gas) assumed in simulations while global properties such as stellar and gas masses and SFRs are almost unchanged. \citet{dcs:20} analyse by post-processing various cosmological simulations with the same method to calculate $L'_{\rm CO(1-0)}$ and $M_{\rm H_2}$. They find that the results do not converge between different sets of simulations. \citet{hsc:20} post-process the IllustrisTNG and original Illustris simulations to reproduce number counts of submillimetre galaxies (SMGs) at the redshift $z=2$. They find that utilising IllustrisTNG significantly underpredicts the number of bright SMGs although their result using the original Illustris simulation is consistent with observations. They argue that this is because galaxies of IllustrisTNG have lower dust masses (metallicities) and SFRs than those of the original Illustris. Bright SMGs generally have quite high SFRs in observations, and this fact implies that such SMGs are expected molecular-rich. Our results predicting the low CO luminosities may stem from the same reason.

\section{Conclusions and summary}
\label{conclusions}
We utilise the IllustrisTNG simulation and populate unresolved gas clouds whose sizes are approximated as `thermal' Jeans lengths of CNM. Adopting DESPOTIC, our method can compute not only CO(1-0) but also higher-$J$ lines, calculate \H2 mass.

For galaxies at the redshift $z=0$, we can reproduce the LF of CO(1-0) obtained by xCOLD GASS. Our values of $\alpha_{\rm CO}$ are consistent with those estimated for the observed galaxies. Although we assume the simple model to approximate the radiation fields, the distribution of model $\alpha_{\rm CO}$ is in agreement with the results of the more detailed model of \citet{nko:12}. We find that about ten per cent of \H2 in the Universe resides in galaxies with $M_{\rm star}\lesssim10^9~{\rm M_\odot}$. These dwarfs have significantly low molecular abundances, which may be a reason for their low SF efficiencies. 

For high-redshift galaxies, our method underpredicts CO luminosities. We find that we can mitigate the discrepancy of the CO-LFs if we enlarge the cloud sizes by a factor of ten, which corresponds to assuming a larger $\lambda_{\rm J}$ than the `thermal' Jeans lengths in high-redshift galaxies. Highly turbulent states of galaxies observed in high-redshift Universe is expected to suppress gravitational collapse in small scales and lead to the formation of large and massive clouds. Thus, our results imply the redshift-evolution of ISM properties in molecular-rich and star-forming galaxies.

Our method enables direct comparison between simulations and observations. It forms the basis for a wealth of future studies, including mock observations using simulations and evaluation for potential biases in kinematic analyses using gas densities and velocity dispersion estimated from measurements of CO intensities and line widths. 

\section*{Acknowledgements}
This study was supported by World Premier International Research Center Initiative (WPI), NAOJ ALMA Scientific Research Grant Number 2019-11A, MEXT, Japan and by SPPEXA through JST CREST JPMHCR1414. SI receives the funding from KAKENHI Grant-in-Aid for Young Scientists (B), No. 17K17677, and HY receives the funding from Grant-in-Aid for Scientific Research (No. 17H04827 and 20H04724) from the Japan Society for the Promotion of Science (JSPS). The numerical computations presented in this paper were carried out on the analysis servers, the general-purpose PC cluster and Cray XC50 at Center for Computational Astrophysics, National Astronomical Observatory of Japan.

\section*{Data availability}
The data underlying this article will be shared on reasonable request to the corresponding author.



\bibliographystyle{mnras}



%




\bsp	
\label{lastpage}
\end{document}